%% file: main.tex
\documentclass[11pt]{article}
\usepackage[utf8]{inputenc}
\usepackage{tikz}
\usepackage{tikz-network}

\usepackage{amssymb}

\usetikzlibrary {arrows.meta,bending,positioning}

\usepackage{authblk} % authors and affiliations formatting
\usepackage{hyperref} % email url

\title{Diagrammatic Rules for Triad Census}

\author[1*]{Enrico Borriello} 
\usepackage{url}

\affil[1]{School of Complex Adaptive Systems, Arizona State University, Tempe, AZ, USA} 
\affil[*]{\small\url{enrico.borriello@asu.edu}}

\date{}

\input{graphs.tex}

\begin{document}

\maketitle

\section*{Abstract}
In network theory, a triad census is a method designed to categorize and enumerate the various types of subgraphs with three nodes and their connecting edges within a network. Triads serve as fundamental building blocks for comprehending the structure and dynamics of networks, and the triad census offers a systematic approach to their classification. Typically, triad counts are obtained numerically, but lesser-known methods have been developed to precisely evaluate them without the need for sampling. In our study, we build upon Moody's matrix approach, presenting general diagrammatic rules that systematically and intuitively generate closed formulas for the occurrence numbers of triads in a network. 

\section{Introduction}

One of the most prolific strategies in the study of a complex system involves the analysis of its network topology --the graph encoding information about the connectivity among its interacting components.

Network theory has experienced an enormous increase in adoption by the scientific community over the past decades. Various complementary strategies have emerged and have been successfully employed to understand the topology of real-world networks \cite{molontay2019two}. On the one hand, global studies have unveiled the connection between the evolutionary nature of real-world networks and the imprint this leaves on their topology; Most notably, the ubiquitous presence of scale-free networks \cite{barabasi1999emergence,strogatz2001exploring,albert2002statistical}. On the other hand, exploring the local topology of complex networks has proven equally fruitful. In this case, a common and systematic strategy involves comparing the frequency of recurrent motifs in complex networks with their occurrence in random counterparts --networks sharing similar global characteristics but with erased peculiarities of the real-world networks they represent.
This motif census has demonstrated particular success when applied to triads \cite{holland1974statistical, holland1977method, fershtman1985transitivity, snijders1986extensions, wasserman1975random}, specifically the smallest non-trivial network subsets consisting of just three nodes. 

Notably, the census of fully connected, directed triads has uncovered a limited number of recurring patterns capable of categorizing extensive ensembles of real-world networks into significant {\it superfamilies} distinguished by distinctive and frequently occurring profiles \cite{milo2002network,milo2004superfamilies}. Arising from its roots in social science \cite{holland1976local}, the triad census has now become a widely utilized and prolific tool for analyzing biological networks, particularly gene regulatory networks \cite{milo2002network, shen2002network, alon2007network, ciriello2008review,borriello2023local}, as well as technology networks such as the World Wide Web (WWW) \cite{zvereva2016triad}. Additionally, it has found applications in linguistics \cite{allen1995natural, vollrath2001proc}.

Historically, the triad census has primarily been conducted numerically, likely due to an initial misconception about the feasibility of deriving closed formulas to count triads \cite{wasserman1994social, wasserman1996logit}. These formulas were eventually provided by Moody \cite{moody1998matrix}. Nonetheless, to this date, they are still criminally underutilized.

The advantage of employing closed formulas, instead of an exhausting scan searching for all triads in a network, not only resides in higher speed across a wide range of network sizes where both algorithms can operate effectively but also in the potential to reveal conservation rules, akin to those identified in \cite{milo2004superfamilies}. Discovering such conservation rules would be challenging, if not impossible, through network-specific numerical scans.

In this work, our goal is to revisit Moody's approach and illustrate how his rules can be consolidated into a general matrix formula. The matrices in this formula can be linked to the three edges of a triad in a manner that enables the provision of just a few diagrammatic rules. These rules correctly allow the reproduction of expressions for all thirteen fully-connected triads.

In the first part of the next section, we will re-derive matrix expressions for the triad census that are equivalent to Moody's expression, but by following a more consistent strategy across various triads. This will lead, in the second half of the section, to the generalization of our matrix expressions through the consistent application of just a few diagrammatic rules. We will then conclude by explicitly demonstrating how triad-specific counts can be deduced directly through the applications of our rules.

\begin{figure}
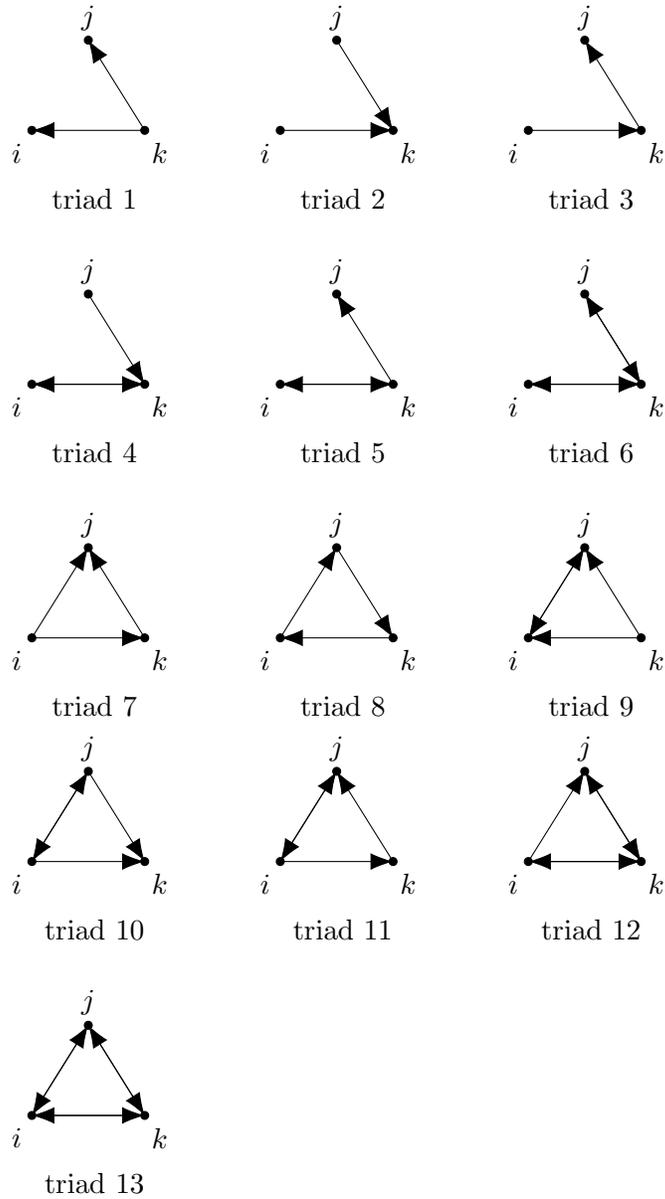

    \centering
\begin{center}    
  \begin{tabular}{ccccc}
    \triadi{1} & \quad & \triadii{1} & \quad & \triadiii{1} \\
    triad 1 & \quad & triad 2 & \quad & triad 3 \\[4mm]
    \triadiv{1} & \quad & \triadv{1} & \quad & \triadvi{1} \\
    triad 4 & \quad & triad 5 & \quad & triad 6 \\[4mm]
    \triadvii{1} & \quad & \triadviii{1} & \quad & \triadix{1} \\
    triad 7 & \quad & triad 8 & \quad & triad 9 \\
    \triadx{1} & \quad & \triadxi{1} & \quad & \triadxii{1} \\
    triad 10 & \quad & triad 11 & \quad & triad 12 \\[4mm]
    \triadxiii{1} & \quad &   & \quad &   \\
    triad 13 & \quad &   & \quad &   
  \end{tabular}
  \label{fig:triads}
  \caption{There are 16 possible types of triads, each representing a unique arrangement of relationships between three nodes in a network. In this work we focus on the thirteen fully connected ones.}
\end{center}
\end{figure}

\section{Methods and Results}

\subsection*{\centering \it Matrix Formulas}

As in \cite{moody1998matrix}, we focus on unweighted networks with no self-loops. Our consideration is limited to fully connected triads, numbered from 1 to 13 as shown in Figure \ref{fig:triads}. For each triad $i$ (where $i=1,\dots,13$), we aim to evaluate its occurrence number $t_i$ in a closed form that avoids explicit references to iterations over combinations of nodes. Instead, all our results are expressed in terms of matrix operations on $A$, the adjacency matrix of the network. Presented as straightforward transformations of $A$, these expressions can be efficiently computed using various optimized packages for vector calculus. (As an example, we complement these notes with a Python implementation that uses NumPy.)

By standardizing the derivative process of our final matrix expression, we aim to highlight the generalization discussed in the next section. In that section, all formulas will be reinterpreted as outcomes derived from the application of common diagrammatic rules.

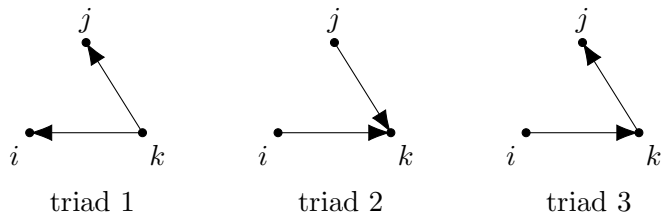
\begin{center}    
  \begin{tabular}{ccccc}
    \triadi{1} & \quad & \triadii{1} & \quad & \triadiii{1} \\
    triad 1 & \quad & triad 2 & \quad & triad 3
  \end{tabular}
\end{center}

\noindent
{\bf Triads 1, 2, and 3:} Given the convention on the vertex labels $i,j$ and $k$ as in figure, triad 1 occurs whenever the elements $A_{pq}$ of the adjacency matrix $A$ of the network satisfy the conditions 
\begin{eqnarray*}
A_{ki} = A_{kj} =1 \ , \\
A_{jk} = A_{ik} = A_{ij} = A_{ji} = 0 \ , \\
i\neq j \ . 
\end{eqnarray*} 

We do not need to impose $k \neq i$ and $k \neq j$ because we will not account  for the presence of self-loops in this manuscript. Therefore, we can just assume $A_{ii}=A_{jj}=0$.

In a network with $n$ nodes, and for given nodes $i$ and $j$, we can calculate the number of nodes like node $k$ in figure:
\begin{equation}
t_{1,ij} = 
(1-A_{ij})(1-A_{ji})
\sum_{k=1}^n (1-A_{jk})(1-A_{ik})A_{ki}A_{kj} \ .
\label{eq:sum1}
\end{equation}

We can then calculate $t_1$, the occurrence number of triad 1 in the network, by summing over all the elements $t_{1,ij}$ with $i \neq j$, and divide the result by 2 to prevent the double-counting induced by the symmetry of this triad with respect to $i$ and $j$:

\begin{equation}
t_1 = \frac{1}{2} \sum_{i,j\neq i} t_{1,ij} \ .
\label{eq:sum2}
\end{equation}

By defining 
\begin{equation}
X = ({\bf 1}-A)\circ({\bf 1}-A^T)
\label{eqX}
\end{equation}
and
\begin{equation}
Y = A \circ A^T \ ,
\label{eqY}
\end{equation}
where ${\bf 1}_{ij}=1\ , \forall i,j$,
and
\begin{equation}
D_1 = A^T  A - Y A
-A^T Y + Y^2 \ ,
\label{D1}
\end{equation} 
we can recast $t_{1,ij}$ as $X_{ij}D_{1,ij}$, and

\begin{equation}
\boxed{
t_1= \frac{1}{2}[\textrm{sum}( X\circ D_1)-\textrm{tr}(X\circ D_1)]
} \ .
\label{T1}
\end{equation}

In the previous formulae: 
$\circ$ is the
the element-wise product between
matrices (Hadamard product);
$\textrm{sum}(\cdot)$ returns the sum of all the element of a matrix; $\textrm{tr}(\cdot)$ is the trace. We have also made use of the fact that $Y = A\circ A^T$ is a symmetric matrix. 

\vspace{4mm}\noindent
We can calculate $t_2$, the occurrence number of triad 2, by a similar argument. As triad 1 becomes triad 2 once $A$ is transposed, we just need to replace $A$ with $A^T$ in (\ref{T1}): $t_2= \left.t_1\right|_{A\rightarrow A^T}$
\begin{equation}
\boxed{
t_1= \frac{1}{2}[\textrm{sum}( X\circ D_2)-\textrm{tr}(X\circ D_2)]
}\ ,
\label{T2}
\end{equation}
with
\begin{equation}
D_2 = A  A^T - Y A^T
-A Y + Y^2 \ ,
\label{D2}
\end{equation}

\vspace{4mm}\noindent
We can count the occurrence number of triad 3 analogously to what we have done with triads 1 and 2. The new conditions being 
\begin{eqnarray*}
A_{ik} = A_{kj} =1 \ , \\
A_{jk} = A_{ki} = A_{ij} = A_{ji} = 0 \ , \\
i\neq j \ . 
\end{eqnarray*} 
We first define 
\begin{equation}
D_3 = A^2  - Y A -A Y + Y^2 \ ,
\label{D3}
\end{equation}
and then cast the count in the form
\begin{equation}
\boxed{
t_3= \textrm{sum}(X\circ D_3)-\textrm{tr}(X\circ D_3)
}\quad .
\label{T3}
\end{equation}
Notice the lack of the factor $1/2$ in this count, as this triad is not symmetric with respect to an exchange of nodes $i$ and $j$. Also notice the common functional form of Equations \ref{T1}, \ref{T2}, and \ref{T3}.

% ----------------------------------------------

\vspace{4mm}\noindent
{\bf Triads 9, 10, and 11:} Let us evaluate the counts for triads 9, 10, and 11 first, because of their similarities with the expressions we have already obtained for triads 1, 2, and 3.

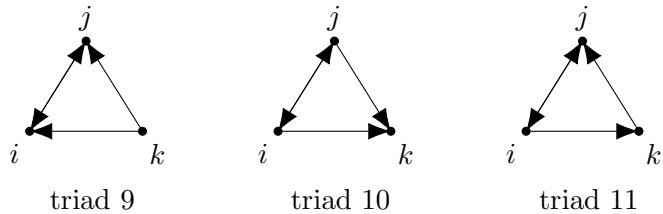
\begin{center}    
  \begin{tabular}{ccccc}
    \triadix{1} & \quad & \triadx{1} & \quad & \triadxi{1} \\
    triad 9  & \quad & triad 10 & \quad & triad 11
  \end{tabular}
\end{center}

The difference this time is the presence of a symmetric connection between nodes $i$ and $j$, translating into factors $A_{ij} = A_{ji} = 1$ as opposed to zero. The result for triad 9 then becomes

\begin{equation}
\boxed{
t_9= \frac{1}{2} \textrm{sum}( Y\circ D_9)
}\ ,
\label{T9}
\end{equation}
Where it is simply $D_9=D_1$ (Equation \ref{D1}), and we are omitting the term $-\textrm{tr}(Y\circ B)$ because $Y_{ii}=0, \forall i$ as we are assuming the absence of self-loops in the networks. (Notice that, as long as the presence of self-loops is not accounted for, we can always replace the diagonal elements $A_{ii}$ of the adjacency matrix with null entries before performing this census.)

Just like we derived $t_2$ from $t_1$, we can now get $t_{10}$ by just replacing $A$ with $A^T$ in Equation (\ref{T9}),
$
t_{10}= \left.t_9\right|_{A\rightarrow A^T}
$, and obtain
\begin{equation}
\boxed{
t_{10}= \frac{1}{2} \textrm{sum}( Y\circ D_{10})
}\ ,
\label{T10}
\end{equation}
where this time it is $D_{10}=D_2$ (Equation \ref{D2}).

It should be clear at thiss point, and it will look more systematica in the next section, that we can easily get $t_{11}$ from our expression for $t_3$ after replacing $X$ with $Y$, and omitting to subract the trace:

\begin{equation}
\boxed{
t_{11}= \textrm{sum}(Y\circ D_{11})
}\ ,
\label{T11}
\end{equation}
where $D_{11} = D_3$ (Equation \ref{D3}).

Not surprisingly, this time, $t_9$, $t_{10}$, and $t_{11}$ once again exhibit the same functional form as the counts we obtained earlier. The only (apparent) distinction is the absence of the trace term, as it is evaluated over matrices with null diagonal entries. Inherited from our previous expression, the numeric coefficient continues to convey information about the symmetry of the triad concerning a swap between indices $i$ and $j$.

% ----------------------------------------------

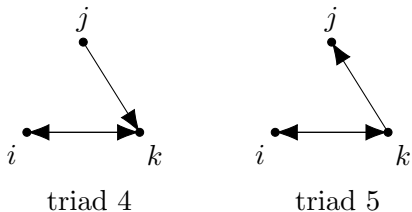
\begin{center}
  \begin{tabular}{ccc}
    \triadiv{1} & \quad & \triadv{1} \\
    triad 4  & \quad & triad 5
  \end{tabular}
\end{center}

\vspace{4mm}\noindent
{\bf Triads 4 and 5:} There is not much novelty involved in the deduction of the occurrence count of triads 4 and 5. For triad 4, we can directly express the number of nodes such as $k$ in our graph as:
\[
t_{4,ij} = X_{ij} \sum_k Y_{ij}A_{jk}({\bf 1}-A_{kj})
\]
and then derive
\[
t_4 = \sum_{i,j\neq i} t_{4,ij} \ .
\]
After defining
\[
D_4 = Y A^T - Y^2 \ , % [D]
\]
we can express our result in a closed form as
\begin{equation}
    \boxed{
t_4 = \textrm{sum}(X\circ D_4) - \textrm{tr}(X\circ D_4)
}\ .
\label{T4}
\end{equation}

\vspace{4mm}

\noindent
Triads of type 5 can be counted in a similar way, with
\[
D_5 = Y A - Y^2 % [E]
\]
and
\begin{equation}
    \boxed{
t_5 = \textrm{sum}(X\circ D_5) - \textrm{tr}(X\circ D_5)
}\ .
\label{T5}
\end{equation}

\vspace{4mm}\noindent
{\bf Triad 6:} We will determine the occurrence number of Triad 6 using a different approach. Our primary objective is to demonstrate that, even with this alternative method, we will arrive at a triad count that shares the same functional form as the expressions obtained thus far.

\begin{center}
  \begin{tabular}{ccc}
     & \quad & \triadvi{1} \\
       & \quad & triad 6
  \end{tabular}
\end{center}

\noindent
Whenever feasible, we will leverage the established property of the powers of the adjacency matrix of an unweighted network to determine the number of paths of a given length between pairs of nodes. For instance, the element $(i, j)$ of $A^l$ equals the number of paths of length $l$ from node $i$ to node $j$.

With this in mind, we can determine the occurrence number of triads of type 6 by focusing solely on bidirected paths of length 2 between $i$ and $j$ with $i \neq j$. It is straightforward to observe that we can identify the components of matrix $A$ with a non-zero symmetric component by taking the element-wise product of $A$ and its transpose $A^T$, what we previously denoted as matrix $Y$ (\ref{eqY}). Consequently, we can reinterpret $Y$ as the adjacency matrix of the subgraph of $A$ that exclusively includes the bidirected edges. Thus, the element $(i, j)$ of $Y^2$ represents the count of bidirected paths of length 2 between nodes $i$ and $j$. Accounting for the symmetric nature of triad 6 concerning $i$ and $j$, we can express our result as:

\begin{equation}
    \boxed{
t_6= \frac{1}{2}
[ \textrm{sum}(X\circ Y^2) - \textrm{tr}(X\circ Y^2) ]
}\ .
\label{T6}
\end{equation}

The Hadamard product with $X$ ensures the absence of connections between nodes $i$ and $j$. Additionally, the trace, as customary at this point, reinforces the condition that $i \neq j$. It is important to note that the diagonal elements of $Y^2$ are not automatically zero, hence necessitating the retention of the trace term.

It is a simple exercise to illustrate that this result can be obtained again by starting from more fundamental assumptions about the elements $A_{ij}$, employing some algebra, akin to the approach taken in this section so far. The more crucial aspect is deducing the functional form for the count that we have become accustomed to. (Considering the simplicity of the second factor in the Hadamard product, there did not seem to be a need to define a matrix $D_6=Y^2$, but one could do so to emphasize the formal analogy even further.)

% ----------------------------------------------

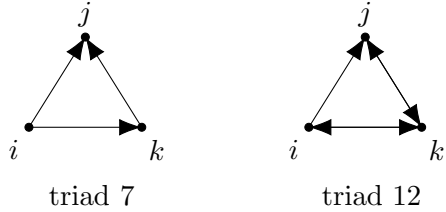
\begin{center}
  \begin{tabular}{ccccc}
    \triadvii{1} &  & \quad & \triadxii{1} \\
    triad 7   &  & \quad & triad 12 
  \end{tabular}
\end{center}

\vspace{4mm}\noindent
{\bf Triads 7 and 12:}

\noindent
At this point, it should not come as a surprise that triad 7 can be counted as triads 3, after we replacement of $X$ with
\[
Z = A \circ (1-A^T) = A -Y
\]
in the final expression for $t_3$
Therefore,
\begin{equation}
\boxed{
t_7 = \textrm{sum}(Z\circ D_7)} \ ,
\label{T7}
\end{equation}
where it is simply $D_7=D_3$.

The same is true about the connection between triads 12 and 6:
\begin{equation}
\boxed{
t_{12} = \textrm{sum}(Z\circ Y^2)} \ .
\label{T12}
\end{equation}

Again, we have ignored the subtraction of traces in (\ref{T7}) and (\ref{T12}) under the assumption that $Z_{ii}=0$ for all $i$, provided we disregard self-loops.

% ----------------------------------------------

\vspace{4mm}\noindent
{\bf Triads 8 and 13:}

\begin{center}    
  \begin{tabular}{ccc}
    \triadviii{1} & \quad & \triadxiii{1} \\
    triad 8    & \quad & triad 13
  \end{tabular}
\end{center}

Similar to triad 6, the counts for triads 8 and 13 possess a geometric significance that is too evident to overlook. Consequently, we will assess them differently.

Given $A_{ii}=0 \ , \forall i$, these triads represent the exclusive ways to create a closed loop of length 3. In the context of triad 13, we additionally necessitate that this loop be bidirectional:

\begin{equation}
\boxed{
t_8 = \frac{1}{3}\textrm{tr}(Z^3)
}\quad 
\label{T8}
\end{equation}

\begin{equation}
\boxed{
t_{13} = \frac{1}{6}\textrm{tr}(Y^3)
}\quad 
\label{T13}
\end{equation}

Interestingly, these expressions exhibit the highest dissimilarity among all we have derived. If they genuinely share the same functional form for all remaining triad counts, this expression implies that the sum terms are null.

This could be demonstrated explicitly, commencing with expressions such as
\[
t_{8,ijk} = A_{ij}A_{jk}A_{ki}
({\bf 1}-A_{ji})
({\bf 1}-A_{kj})
({\bf 1}-A_{ik})
\]
and proceeding with the necessary algebraic steps. It will be important to establish this proof to reconcile these counts with the expression derived from the straightforward application of the diagrammatic rules we are about to present.

\subsection*{\centering \it Diagrammatic Rules}

We can now establish diagrammatic rules for deducing the matrix expression of the triad count directly from its graph. We have seen that, for a given ordered pair of generic nodes $p$ and $q$, the absence of a connection between them corresponds to the presence in the result of a matrix $X$. An outgoing edge (from $p$ to $q$) is associated with a matrix $Z$, an incoming edge ($q$ to $p$) with $Z^T$, and a symmetric edge between the two with a matrix $Y$. The required matrix operations involving the matrices associated to the three edges in a triad can be derived from the two general steps applied in calculating the terms $t_{\alpha}$ $(\alpha=1,\dots,13)$, like in Equations \ref{eq:sum1} and \ref{eq:sum2}. Initially, we sum over the intermediate nodes that, along with $p$ and $q$, satisfy the graph topology of the triad being counted. Subsequently, an additional sum is conducted over nodes $p$ and $q$, taking into account the graph's symmetry. 

The list below outlines these characteristic matrices along with their explicit expressions in terms of $A$ for any arbitrary ordered pair of nodes $p$ and $q$:

\begin{center}    
  \begin{tabular}{lll}
    $p$ \diadZero{1}  $q$  & $C^{[p,q]} = X   $ &  $\equiv \quad (A-Y) \circ (A^T - Y)$  \\
    $p$ \diadItoJ{1}  $q$  & $C^{[p,q]} = Z   $ &  $\equiv \quad A \circ (1-A^T) = A - Y$  \\
    $p$ \diadJtoI{1}  $q$  & $C^{[p,q]} = Z^T $ &  $\equiv \quad A^T \circ (1-A) = A^T - Y$  \\
    $p$ \diadIandJ{1} $q$  & $C^{[p,q]} = Y   $ &  $\equiv \quad A \circ A^T $  \\
  \end{tabular}
\end{center}

With the conventions and labeling established in Figure \ref{fig:triads}, We will refer to the first two nodes we selected, $i$ and $j$ as the {\it basis} of the triad, and to the intermediate node $k$ as the {\it vertex}. 

We first build the matrix
\begin{equation}
\boxed{
B_{\alpha} = C^{[i,j]} \circ \left( C^{[k,i]\,T} \cdot C^{[k,j]}   \right)
} \ ,    
\end{equation}
where the dot product corresponds to the sum over $k$ (e.g. Equation \ref{eq:sum2}), and the Hadamard product instead corresponds to the sum over $i$ and $j$ (e.g. Equation \ref{eq:sum1}).
The final result can then be expressed as
\begin{equation}
\boxed{
t_{\alpha} = \frac{1}{s_{\alpha}} \left[ \textrm{sum}(B_{\alpha})-\textrm{tr}(B_{\alpha}) \right]
} \ ,
\label{eq:general_formula}
\end{equation}
where $s_{\alpha}$ is a symmetry factor, defined as follows.

We will define a triad as ``rotationally-symmetric'' if it remains unchanged under a cyclic permutation of all three indices. A triad that remains unchanged after a label exchange of the nodes in its basis will be said to be ``parity-symmetric.''

According to these definitions:

\begin{itemize}
    \item Triads 3, 4, 5, 7, 11, and 12 do not possess any of these symmetries.
    \item Triads 1, 2, 6, 9, and 10 are parity-symmetric.
    \item Triad 8 is rotationally-symmetric.
    \item Triad 13 is both parity- and rotationally-symmetric.
\end{itemize}

For asymmetric triads, $s_{\alpha} = 1$. For parity-symmetric triads, $s_{\alpha}=2$. For parity- and rotationally-symmetric triads, $s_{\alpha} = 2 \times 3 = 6$.

Now, we need to demonstrate that equation \ref{eq:general_formula} indeed provides the general functional form for all triad counts.
The proofs for triads 1 to 5, 7, and 9 to 12 are already provided explicitly in the previous section by the method we adopted to count their occurrence numbers, leaving us with the task of confirming the correctness of our diagrammatic rules for triads 6, 8, and 13.

In the case of triad 6, the basis of the triad contributes the matrix $C^{[i,j]}=X$, while the vertices correspond to matrices $C^{[k,i]} = C^{[k,j]} = Y$. With $s_6 = 2$, Equations \ref{eq:general_formula} and \ref{T6} actually yield the same result.

The remaining proofs for triads 8 and 13 are similar. We will only present the quicker proof for triad 13, but the interested reader can readily perform the necessary algebra to verify the other one.

According to the diagrammatic rules provided here, $t_{13} = \frac{1}{6}\textrm{sum}(Y\circ Y^2)$, to be compared to $t_{13} = \frac{1}{6}\textrm{tr}(Y^3)$ as obtained from counting closed and bidirectional paths of length 3 in the network. Therefore, we just need to proof that $\textrm{tr} (Y^3) = \textrm{sum} (Y\circ Y^2)$. The element $ij$ of $Y^3$ can be written as
\[
(Y^3)_{ij} = \sum_{q,p} Y_{ip} Y_{pq} Y_{qj} \ . 
\]
Therefore
\[
\textrm{tr} (Y^3) = \sum_i (Y^3)_{ii} = \sum_{i,p,q} Y_{ip} Y_{pq} Y_{qj}  \ .  
\]
On the other hand,
\[
(Y\circ Y^2)_{ij} = Y_{ij} \sum_p Y_{ip} Y_{pj}
\]
and
\[
\textrm{sum} (Y\circ Y^2) = \sum_{i,j,p} Y_{ij} Y_{ip} Y_{pj} \ .
\]
By virtue of the symmetry of Y:
\[
\textrm{sum} (Y\circ Y^2)  = \sum_{i,j,p} Y_{ij} Y_{jp} Y_{pi} 
= \textrm{tr} (Y^3) \ , 
\]
as we wanted to prove.

We can therefore establish the equivalence between the general expression in Equation \ref{eq:general_formula}
and the individual triad counts from the previous subsection.

\section{Conclusions}

In network theory, a triad census is a method aimed at categorizing and enumerating different types of subgraphs with three nodes and their connecting edges within a network. Triads, as fundamental building blocks, play a crucial role in understanding the structure and dynamics of networks. The triad census provides a systematic approach to classify these structures, allowing researchers to analyze the prevalence of various triad types in a given network. This method is particularly valuable in social science for studying local structures and clustering properties of social networks.

Moody is credited with the initial derivation of closed formulas for counting the occurrence number of triads. Despite its significance, the scientific community has been hesitant to fully embrace the potential of deducing triad census through efficient matrix operations. Traditionally, numerical methods have been utilized to count occurrences of each triad type through exhaustive network traversal algorithms. These algorithms, however, only suboptimally leverage the peculiar structure of sparse connectedness present in most real-world networks. 

Our study introduces diagrammatic rules that systematically and intuitively generate closed formulas for the occurrence number of triads in a network. This unifies Moody's matrix expressions into a common generating rule. The simplification presented in this work, characterized by an extremely concise mathematical representation, has the potential to enhance the understanding of underlying patterns and dynamics within a network. This is achieved by directly relating the triad counts to common matrix operations performed over the network's adjacency matrix. Additionally, we envision this work as an initial step towards generalizing diagrammatic rules for the systematic counting of larger and more complex motifs.

\bibliographystyle{unsrt}
\bibliography{bibliography}

\end{document}

%% file: graphs.tex
\newcommand{\triadi}[1]{
    \begin{tikzpicture}[scale=#1]
    \draw[fill=black] (0,0) circle (1.5pt);
    \draw[fill=black] (3/2,0) circle (1.5pt);
    \draw[fill=black] (0.75,1.2) circle (1.5pt);
    \node at (-0.2,-0.3) {$i$};
    \node at (1.7,-0.3) {$k$};
    \node at (0.75,1.5) {$j$};
    \draw[{Latex[length=3mm]}-] (0,0) -- (3/2,0);
    \draw[-{Latex[length=3mm]}] (3/2,0) -- (0.75,1.2);
    \end{tikzpicture} 
}

\newcommand{\triadii}[1]{
    \begin{tikzpicture}[scale=#1]
    \draw[fill=black] (0,0) circle (1.5pt);
    \draw[fill=black] (3/2,0) circle (1.5pt);
    \draw[fill=black] (0.75,1.2) circle (1.5pt);
    \node at (-0.2,-0.3) {$i$};
    \node at (1.7,-0.3) {$k$};
    \node at (0.75,1.5) {$j$};
    \draw[{Latex[length=3mm]}-] (3/2,0) -- (0,0);
    \draw[-{Latex[length=3mm]}] (0.75,1.2) -- (3/2,0) ;
    \end{tikzpicture}
}

\newcommand{\triadiii}[1]{
    \begin{tikzpicture}[scale=#1]
    \draw[fill=black] (0,0) circle (1.5pt);
    \draw[fill=black] (3/2,0) circle (1.5pt);
    \draw[fill=black] (0.75,1.2) circle (1.5pt);
    \node at (-0.2,-0.3) {$i$};
    \node at (1.7,-0.3) {$k$};
    \node at (0.75,1.5) {$j$};
    \draw[{Latex[length=3mm]}-] (3/2,0) -- (0,0);
    \draw[-{Latex[length=3mm]}]  (3/2,0) --(0.75,1.2) ;
    \end{tikzpicture}
}

\newcommand{\triadiv}[1]{
    \begin{tikzpicture}[scale=#1]
    \draw[fill=black] (0,0) circle (1.5pt);
    \draw[fill=black] (3/2,0) circle (1.5pt);
    \draw[fill=black] (0.75,1.2) circle (1.5pt);
    \node at (-0.2,-0.3) {$i$};
    \node at (1.7,-0.3) {$k$};
    \node at (0.75,1.5) {$j$};
    \draw[{Latex[length=3mm]}-] (3/2,0) -- (0,0);
        \draw[{Latex[length=3mm]}-] (0,0) -- (3/2,0);
    \draw[-{Latex[length=3mm]}] (0.75,1.2) -- (3/2,0) ;
    \end{tikzpicture}
}

\newcommand{\triadv}[1]{
    \begin{tikzpicture}[scale=#1]
    \draw[fill=black] (0,0) circle (1.5pt);
    \draw[fill=black] (3/2,0) circle (1.5pt);
    \draw[fill=black] (0.75,1.2) circle (1.5pt);
    \node at (-0.2,-0.3) {$i$};
    \node at (1.7,-0.3) {$k$};
    \node at (0.75,1.5) {$j$};
    \draw[{Latex[length=3mm]}-] (3/2,0) -- (0,0);
    \draw[{Latex[length=3mm]}-] (0,0) -- (3/2,0);
    \draw[-{Latex[length=3mm]}] (3/2,0) -- (0.75,1.2) ;
    \end{tikzpicture}
}

\newcommand{\triadvi}[1]{
    \begin{tikzpicture}[scale=#1]
    \draw[fill=black] (0,0) circle (1.5pt);
    \draw[fill=black] (3/2,0) circle (1.5pt);
    \draw[fill=black] (0.75,1.2) circle (1.5pt);
    \node at (-0.2,-0.3) {$i$};
    \node at (1.7,-0.3) {$k$};
    \node at (0.75,1.5) {$j$};
    \draw[{Latex[length=3mm]}-] (3/2,0) -- (0,0);
    \draw[{Latex[length=3mm]}-] (0,0) -- (3/2,0);
    \draw[-{Latex[length=3mm]}] (3/2,0) -- (0.75,1.2) ;
    \draw[-{Latex[length=3mm]}] (0.75,1.2) -- (3/2,0) ;
    \end{tikzpicture}
}

\newcommand{\triadvii}[1]{
    \begin{tikzpicture}[scale=#1]
    \draw[fill=black] (0,0) circle (1.5pt);
    \draw[fill=black] (3/2,0) circle (1.5pt);
    \draw[fill=black] (0.75,1.2) circle (1.5pt);
    \node at (-0.2,-0.3) {$i$};
    \node at (1.7,-0.3) {$k$};
    \node at (0.75,1.5) {$j$};
    \draw[{Latex[length=3mm]}-] (3/2,0) -- (0,0);
    \draw[-{Latex[length=3mm]}] (3/2,0) --(3/4,1.2) ;
    \draw[-{Latex[length=3mm]}] (0,0) -- (3/4,1.2) ;
    \end{tikzpicture}
}

\newcommand{\triadviii}[1]{
    \begin{tikzpicture}[scale=#1]
    \draw[fill=black] (0,0) circle (1.5pt);
    \draw[fill=black] (3/2,0) circle (1.5pt);
    \draw[fill=black] (0.75,1.2) circle (1.5pt);
    \node at (-0.2,-0.3) {$i$};
    \node at (1.7,-0.3) {$k$};
    \node at (0.75,1.5) {$j$};
    \draw[-{Latex[length=3mm]}] (3/2,0) -- (0,0);
    \draw[{Latex[length=3mm]}-] (3/2,0) --(3/4,1.2) ;
    \draw[{Latex[length=3mm]}-] (3/4,1.2) --(0,0) ;
    \end{tikzpicture}
}

\newcommand{\triadix}[1]{
    \begin{tikzpicture}[scale=#1]
    \draw[fill=black] (0,0) circle (1.5pt);
    \draw[fill=black] (3/2,0) circle (1.5pt);
    \draw[fill=black] (0.75,1.2) circle (1.5pt);
    \node at (-0.2,-0.3) {$i$};
    \node at (1.7,-0.3) {$k$};
    \node at (0.75,1.5) {$j$};
    \draw[{Latex[length=3mm]}-] (0,0) -- (3/2,0);
    \draw[-{Latex[length=3mm]}] (3/2,0) -- (3/4,1.2) ;
    \draw[-{Latex[length=3mm]}] (3/4,1.2) --(0,0) ;
    \draw[-{Latex[length=3mm]}] (0,0) -- (3/4,1.2) ;
    \end{tikzpicture}
}

\newcommand{\triadx}[1]{
    \begin{tikzpicture}[scale=#1]
    \draw[fill=black] (0,0) circle (1.5pt);
    \draw[fill=black] (3/2,0) circle (1.5pt);
    \draw[fill=black] (0.75,1.2) circle (1.5pt);
    \node at (-0.2,-0.3) {$i$};
    \node at (1.7,-0.3) {$k$};
    \node at (0.75,1.5) {$j$};
    \draw[{Latex[length=3mm]}-] (3/2,0) -- (0,0);
    \draw[-{Latex[length=3mm]}] (3/4,1.2) -- (3/2,0) ;
    \draw[-{Latex[length=3mm]}] (3/4,1.2) --(0,0) ;
    \draw[-{Latex[length=3mm]}] (0,0) -- (3/4,1.2) ;
    \end{tikzpicture}
}

\newcommand{\triadxi}[1]{
    \begin{tikzpicture}[scale=#1]
    \draw[fill=black] (0,0) circle (1.5pt);
    \draw[fill=black] (3/2,0) circle (1.5pt);
    \draw[fill=black] (0.75,1.2) circle (1.5pt);
    \node at (-0.2,-0.3) {$i$};
    \node at (1.7,-0.3) {$k$};
    \node at (0.75,1.5) {$j$};
    \draw[{Latex[length=3mm]}-] (3/2,0) -- (0,0);
    \draw[-{Latex[length=3mm]}] (3/2,0) --(3/4,1.2) ;
    \draw[-{Latex[length=3mm]}] (0,0) -- (3/4,1.2) ;
        \draw[-{Latex[length=3mm]}] (3/4,1.2)  -- (0,0);
    \end{tikzpicture}
}

\newcommand{\triadxii}[1]{
    \begin{tikzpicture}[scale=#1]
    \draw[fill=black] (0,0) circle (1.5pt);
    \draw[fill=black] (3/2,0) circle (1.5pt);
    \draw[fill=black] (0.75,1.2) circle (1.5pt);
    \node at (-0.2,-0.3) {$i$};
    \node at (1.7,-0.3) {$k$};
    \node at (0.75,1.5) {$j$};
    \draw[{Latex[length=3mm]}-] (3/2,0) -- (0,0);
    \draw[{Latex[length=3mm]}-] (0,0) -- (3/2,0) ;
    \draw[-{Latex[length=3mm]}] (3/2,0) --(3/4,1.2) ;
    \draw[-{Latex[length=3mm]}] (3/4,1.2) -- (3/2,0) ;
    \draw[-{Latex[length=3mm]}] (0,0) -- (3/4,1.2) ;
    \end{tikzpicture}
}

\newcommand{\triadxiii}[1]{
    \begin{tikzpicture}[scale=#1]
    \draw[fill=black] (0,0) circle (1.5pt);
    \draw[fill=black] (3/2,0) circle (1.5pt);
    \draw[fill=black] (0.75,1.2) circle (1.5pt);
    \node at (-0.2,-0.3) {$i$};
    \node at (1.7,-0.3) {$k$};
    \node at (0.75,1.5) {$j$};
    \draw[{Latex[length=3mm]}-] (3/2,0) -- (0,0);
    \draw[{Latex[length=3mm]}-] (0,0) -- (3/2,0) ;
    \draw[-{Latex[length=3mm]}] (3/2,0) --(3/4,1.2) ;
    \draw[-{Latex[length=3mm]}] (3/4,1.2) -- (3/2,0) ;
    \draw[-{Latex[length=3mm]}] (0,0) -- (3/4,1.2) ;
    \draw[-{Latex[length=3mm]}] (3/4,1.2) -- (0,0);
    \end{tikzpicture}
}

\newcommand{\diadZero}[1]{
    \begin{tikzpicture}[scale=0.5*#1]
    \draw[fill=black] (0,0) circle (1.5pt);
    \draw[fill=black] (3/2,0) circle (1.5pt);
    \end{tikzpicture} 
}

\newcommand{\diadItoJ}[1]{
    \begin{tikzpicture}[scale=0.5*#1]
    \draw[fill=black] (0,0) circle (1.5pt);
    \draw[fill=black] (3/2,0) circle (1.5pt);
    \draw[{Latex[length=1.5mm, width=1mm]}-] (3/2,0) -- (0,0);
    \end{tikzpicture} 
}

\newcommand{\diadJtoI}[1]{
    \begin{tikzpicture}[scale=0.5*#1]
    \draw[fill=black] (0,0) circle (1.5pt);
    \draw[fill=black] (3/2,0) circle (1.5pt);
    \draw[{Latex[length=1.5mm, width=1mm]}-] (0,0) -- (3/2,0);
    \end{tikzpicture} 
}

\newcommand{\diadIandJ}[1]{
    \begin{tikzpicture}[scale=0.5*#1]
    \draw[fill=black] (0,0) circle (1.5pt);
    \draw[fill=black] (3/2,0) circle (1.5pt);
    \draw[{Latex[length=1.5mm, width=1mm]}-] (3/2,0) -- (0,0);
    \draw[{Latex[length=1.5mm, width=1mm]}-] (0,0) -- (3/2,0);
    \end{tikzpicture} 
}